\documentclass{article}
\usepackage{url}

\usepackage[english]{babel}

\usepackage[letterpaper,top=2cm,bottom=2cm,left=3cm,right=3cm,marginparwidth=1.75cm]{geometry}

\usepackage{amsmath}
\usepackage{amssymb}
\usepackage{bm}
\usepackage{graphicx}
\usepackage{hyperref}
\hypersetup{
  bookmarks=false,
  colorlinks=true,
  linkcolor=blue,
  citecolor=blue,
  urlcolor=blue
}
\usepackage{subcaption}
\usepackage{authblk}

\begin{document}

\title{The Shocking Origin of the Flat $EE/BB$ Ratio}

\author[2]{Raphael Flauger}
\author[2]{Alexei G. Kritsuk}
\author[1,2]{Guanhao Sun}

\affil[1]{Department of Physics and Chongqing Key Laboratory for Strongly Coupled Physics, Chongqing University, Chongqing 401331, People’s Republic of China}
\affil[2]{Department of Physics, University of California, San Diego,  La Jolla, CA 92093-0319, USA}

\date{}

\maketitle

\begin{abstract}
Polarized emission from dust and synchrotron radiation from the ISM are the dominant foregrounds for CMB polarization and are a major challenge for extracting the primordial signal on large angular scales. A key characteristic of the galactic foreground emission is its $EE/BB$ ratio. We argue that MHD shocks play an important role in setting the observed $EE/BB$ ratio. To support this, we first analyze quasi-linear magnetohydrodynamics (MHD) simulations to obtain an $EE/BB$ ratio that increases as $\sim k^2$, then show that with increasing energy injection rates, the $EE/BB$ ratio flattens to a value $\gtrsim 1$, approaching observational results. Looking at the distribution of the velocity divergence, a tail with power law $-7/2$ develops around the same injection rates where the $EE/BB$ ratio flattens. While the system becomes more isotropic, MHD shocks are intrinsically anisotropic and lead to the $E/B$ power asymmetry. We also observe total pressure balance among all our simulations, indicating slow wave dominance. Therefore, in the regime we consider, it is important to go beyond linear MHD equations to understand the foreground radiation.
\end{abstract}

\section{Introduction}
The polarization of the CMB contains invaluable information about the primordial evolution of our universe. Scalar perturbations generate $E$-mode polarization. On small angular scales, the $E$-mode polarization contains key information about primordial physics, such as the effective number of relativistic species, the sum of neutrino masses, and the primordial helium abundance. On large angular scales, the $E$-mode polarization is a powerful probe of the epoch of reionization. On small scales, gravitational lensing of the CMB by large scale structure converts the $E$-mode polarization from scalar perturbations into $B$-mode polarization. On large angular scales, $B$-mode polarization is a probe of primordial gravitational waves generated during inflation~\cite{Kamionkowski:2015yta}. 

However, Galactic foreground emission dominates the imprint of primordial gravitational waves at all frequencies even in the cleanest patches of the sky~\cite{Dickinson:2016xyz}. Therefore, a significant amount of effort has been put into understanding the nature of foreground emission in order to eventually separate out the primordial signal. At low frequencies, the dominant source of foreground is the polarized synchrotron radiation from relativistic electrons spiraling around magnetic fields. At high frequencies, the dominant source of foreground is the polarized emission from interstellar dust. One of the key observables for the foreground is the ratio between the power in $E$- and $B$-modes, which holds important information about the underlying physics of the ISM. 

There are several studies that utilize MHD  models of the turbulent ISM to investigate this problem. For example, work in~\cite{Caldwell:2016xkd,10.1093/mnrasl/slx128,Kandel:2017xjx}
studied the contributions of various MHD modes to the observed polarization spectra and debated whether or not we can understand the ratio through such linear mode analysis. Reference \cite{Stalpes:2024lwp} showed that sufficiently strong supersonic turbulence could reproduce the desired ratio, and in~\cite{Kritsuk:2017aab} the authors argued that MHD simulations of multiphase turbulence complemented with a simple dust model can yield the observed result. Recent work by \cite{Ho:2024wls} supported these claims, using higher-resolution simulations of magnetized multiphase turbulence, which successfully reproduced a number of Galactic dust polarization observables, including the $EE/BB$ ratio, refining our understanding of the origin of polarized foreground emissions. 

In this work, we focus on the simplified situation of single-phase isothermal turbulence simulations with varying energy injection rates. We first consolidate our analytical understanding at the quasi-linear low injection rate regime and then observe a ``phase transition" when shocks start to form with increasing energy injection rates. The behavior of the spectra of the MHD variables and the synthetic polarization maps show significant qualitative differences between the low and high energy injection rate regimes, with high injection rate more akin to observational results. We therefore argue that shocks in MHD are important for understanding the observed polarization from thermal dust emission.

\section{Numerical model} 
We used a simplified setup for forced homogeneous compressible MHD turbulence simulations in a periodic domain with a mean magnetic field. The system evolved from uniform, static magnetized initial conditions to a statistically stationary turbulent state in a near-isothermal fluid with an ideal gas equation of state $p=(\gamma-1)\rho e$. Here $p$ is the pressure, $\rho$ -- fluid density, $e$ -- specific internal energy per unit mass, and the ratio of specific heats $\gamma$ is set close to unity, $\gamma=1.002$. We use a standard Ornstein-Uhlenbeck forcing implementation with purely solenoidal large-scale acceleration and autocorrelation time that corresponds to the large-eddy turnover time of the turbulent system to resupply the kinetic energy with an injection rate $\epsilon$. The model includes explicit viscosity $\nu$ and magnetic diffusivity $\eta=\nu$, ensuing small-scale energy dissipation channels $\varepsilon_{\rm kin}=\nu\langle\vec\omega^2+\frac{4}{3}\theta^2\rangle$ and $\varepsilon_{\rm mag}=\eta\langle\vec j^2\rangle$, where $\vec\omega=\vec\nabla\times\vec v$, $\theta\equiv\vec\nabla\cdot\vec v$, $\vec j=\vec\nabla\times\vec B$, $\vec v$ and $\vec B$ are the velocity and magnetic field vectors, and angular brackets $\langle .\rangle$ indicate spatio-temporal average. In statistically stationary homogeneous turbulence large-scale energy injection is matched by small-scale energy dissipation ($\epsilon=\varepsilon_{\rm kin}+\varepsilon_{\rm mag}$).

We set our units so that lengths are measured in terms of the box size, gas densities are normalized by the initial gas density, and the speed of sound is $c_{\rm s}\equiv\sqrt{\gamma p_0/\rho_0}\approx1$. Magnetic fields are measured in units of the mean field strength, so that the Alfv\'en speed $v_{\rm A}\equiv B_0/\sqrt{\mu_0\rho_0}=1$. The set of simulations discussed below is parametrized by the kinetic energy injection rate $\epsilon\in\{0.001,0.010, 0.016, 0.025, 0.040, 0.063,\\ 0.100\}$ and $\nu=\eta=10^{-3}$ and $3.3\times 10^{-4}$ at grid resolutions $256^3$ and $512^3$, respectively. We evolved each case for at least 20 dynamical times, $\tau_{\rm d}\equiv L/2\sqrt{\langle u^2\rangle}$, to generate data for fully developed MHD turbulence. Sonic turbulence Mach numbers of resulting stationary turbulence are as follows $M_{\rm t}\in\{0.16,0.36,0.45,0.52,0.62,0.67,0.81\}$. In compressible hydrodynamic turbulence, at $M_{\rm t}<0.3$ the flow is essentially incompressible, then eddy shocklets (small, localized shock-like structures) start to appear at $M_{\rm t}\in[0.3,0.6]$, then at $M_{\rm t}>0.6$ shocks (strong local compression structures) are generated by turbulent eddies \cite{lee..91,wang..17}.

The equations are solved numerically, using a high order accurate finite difference nonlinear filter method with adaptive dissipation control \cite{yee.07,Yee-Sjogreen-ICOSAHOM09,yee.24}. An ordinary differential equation solver is used to advance the forcing source terms using Strang operator splitting \cite{strang68}. Overall, the method solving the ideal compressible MHD equations is 7th-order accurate in space and uses 3rd-order accurate Runge-Kutta time integration.
The base central scheme is 8th-order accurate in space, uses the Ducros et al. skew-symmetric split form of the inviscid flux derivatives \cite{ducros.....00} for the gas dynamics part of the equations, and avoids splitting of the magnetic field terms to preserve the $\vec\nabla\cdot\vec B=0$ condition. The method uses a wavelet sensor \cite{Sjogreen-Yee-wave-2004} and a nonlinear filter with a  $\vec\nabla\cdot\vec B$-preserving 7th-order weighted essentially non-oscillatory (WENO7) shock-capturing scheme analogous to the scheme described in \cite{seo.23}.

\section{Quasi-linear regime}
At lowest energy injection rate, we have good analytic control over the simulation. The system is quasi-linear in the sense that most non-linear terms are not important, and we can approximate the system with minimal modifications to the simplest linear equations of motion of MHD. We will show this by explicit comparisons between calculated and measured quantities for different kinds of fluctuations, setting up a framework for understanding the system in the quasi-linear regime. 

Mode decomposition from the data shows that in our simulations the Alfv\'en and slow magnetosonic modes dominate, so we will focus on these two modes and ignore fast mode for the rest of the work. 

\subsection{Velocity, density, and magnetic fluctuations}
From the linearized MHD equations one can easily obtain relations between the velocity, density, and magnetic field fluctuations for different modes, and express the density and magnetic fluctuations as functions of the velocity fluctuations~\cite{Caldwell:2016xkd,Kandel:2017xjx}. We test these relations by predicting fluctuations in the density and magnetic fields from measured velocity field, and compare them with direct measurements. 

Before the actual data analysis, we note two subtleties for applying these equations. First, we emphasize that the advection terms, where $\vec v \cdot \vec \nabla$ is applied to the fields, play a non-negligible role in this analysis. This is the major non-linear effect in this regime, hence the name `quasi-linear.' One can then recover the dominating non-linear effect by including only the longest modes of the velocity field, $\vec v \cdot \vec \nabla \sim \vec v_{\rm long} \cdot \vec \nabla$. For simplicity we will write $\vec v_{\rm long} = \vec u$ and treat it as a given background field. In our analysis, we use the $k=0$ and $k=k_{\rm min}$ modes to construct $\vec v_{\rm long}$. The effect of including such terms can be taken into account by first replacing the frequencies $\omega$ with a modified value, 
\begin{equation}
    \omega \rightarrow \bar\omega = \omega + \vec u (\vec x) \cdot \vec k \ \ ,
\end{equation}
then replacing the partial time derivative with total time derivative\footnote{In deriving the dispersion relations, we will be taking two total derivatives. However, the total derivative acting on the advection term is subleading compared with other terms so we can safely ignore them and commute the derivatives.}, 
\begin{equation}
    \partial_t \rightarrow \frac{d}{dt} = \partial_t + \vec u(\vec x)\cdot \vec \nabla \ \ .
\end{equation}
Note that although our data have a strong mean magnetic field with small fluctuations, what we do is slightly different from the `nearly incompressible MHD' (NI MHD) treatment in \cite{zank93}. Here we do not separate our variables into compressible and incompressible parts, and the expansion is around the mean values instead of the incompressible values. We make this choice simply for the convenience of our analysis, and it would be interesting to look at the NI MHD behavior of our data.

\begin{figure}
     \centering
     \begin{subfigure}[b]{0.49\textwidth}
         \centering
         \includegraphics[width=\textwidth]{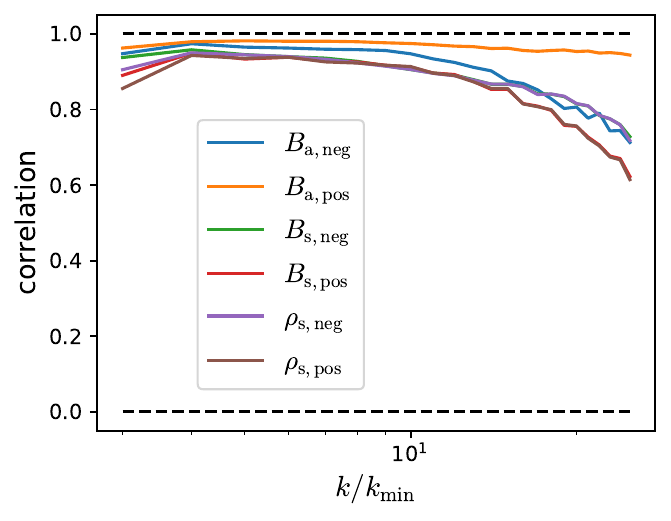}
     \end{subfigure}
     \hfill
     \begin{subfigure}[b]{0.49\textwidth}
         \centering
         \includegraphics[width=\textwidth]{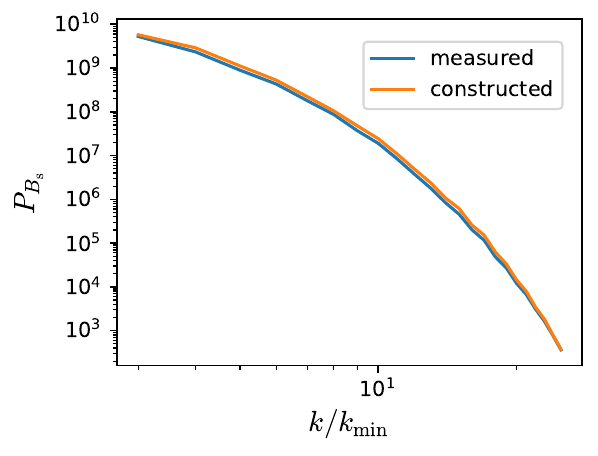}
     \end{subfigure}
     
    \caption{Correlations between measured and calculated fluctuations, and the measured and constructed spectra for $B_{\rm s}$ for injection rate $\epsilon = 0.001$. We see that the linear equations work well in the sense that the correlations between the two sides of the equations of motion are close to 1 for a good range of $k/k_{\rm min}$, and the measured and constructed spectra match within the same range.}
    \label{fig:linear_check}
\end{figure}

After taking into account the advection term, one needs to further separate the positive and negative frequency modes and compare these two components separately. For example, for the magnetic field fluctuation, 
\begin{align}
    \frac{dB}{dt} =& \frac{dB_{\text{pos}}}{dt} +  \frac{dB_{\text{neg}}}{dt} = -i \bar\omega B_{\text{pos}} + i \bar\omega B_{\text{neg}}, \\
    B_{\text{pos}} =& -\frac{1}{2i \bar\omega} \left(\frac{dB}{dt} - i \bar\omega B\right), \\
    B_{\text{neg}} =&\;\;\;\; \frac{1}{2i \bar\omega} \left(\frac{dB}{dt} + i \bar\omega B\right),
\end{align}
where the frequency $\bar\omega$ and derivative $d/dt$ are the ones defined above. Applying these to both magnetic and density fluctuations for both Alfv\'en and slow modes, we arrive at the equations\footnote{A derivation without the positive and negative decomposition and the advection terms can be found, e.g., in \cite{Caldwell:2016xkd}. Here we simply flip the signs between the two sides depending on positive/negative frequencies, and replace $\omega$ with $\bar \omega$ to include advection effects.}
\begin{align}
    \delta \vec B_{\rm a, pos/neg} =& \mp v_{\rm pos/neg} \frac{B_0}{v_{A}} \frac{\hat k \times \hat B_0}{\sin \alpha}, \\
    \delta \vec B_{\rm s, pos/neg} =& \pm v_{\rm pos/neg} \frac{kB_0}{\bar\omega} \frac{\zeta_{\rm s} \sin \alpha}{(\cos^2 \alpha + \zeta_{\rm s}^2 \sin^2 \alpha)^{1/2}} \frac{-\hat k \times (\hat k \times \hat B_0)}{\sin \alpha},\\
    \frac{\delta \rho_{\rm s,pos/neg} }{\rho_0} =& \pm v_{\rm pos/neg} \frac{k}{\bar\omega} \frac{(\cos^2 \alpha + \zeta_{\rm s} \sin^2 \alpha)}{(\cos^2 \alpha + \zeta^2_{\rm s} \sin^2 \alpha)^{1/2}} \ \ ,
\end{align}
with 
\begin{equation}
    \zeta_{\rm s} = \frac{D_{--}}{D_{++}} \cot^2 \alpha , \;\; D = \left(1+\frac{\beta }{2}\right)^2 - 2 \beta \cos^2 \alpha, \;\; D_{\pm \pm} = 1 \pm \sqrt{D} \pm \frac{\beta}{2} \ .
\end{equation}
Here $\beta = 2(c_{s}/v_{A})^2$ is the plasma beta, $\hat B_0$ and $\hat k$ are the unit vectors along the mean magnetic field and the wavevector, respectively, and $\alpha$ is the angle between $\vec B_0$ and $\vec k$. To check these equations, we measure the positive and negative frequency fluctuations on the left hand side directly from the simulations, and calculate the right hand side from the velocity field. We then calculate the correlations between the left and right hand sides, and see that the correlations are decently close to $1$ when $k/k_{\rm min}$ is in the intermediate range, as shown in Fig.~\ref{fig:linear_check}a. The correlations are lower at lowest $k$'s due to energy injection and at highest $k$'s due to significant dissipation effects. We also include the predicted and measured power spectra for the slow modes of the magnetic field fluctuation, as an example to show that the spectra calculations using the linear equations also work within the same region of validity, see Fig.~\ref{fig:linear_check}b. When calculating the spectra, notice that there is a non-zero mixing between the positive and negative frequency parts, and the positive/negative decomposition is important in recovering the expected results. More explicitly, for example, for slow mode fluctuations of the magnetic field, if we use the relation 
\begin{equation}
    \delta B_{\rm s,pos/neg}/B_0 = \pm f_{\rm s} v_{\rm s,pos/neg} \ ,
\end{equation}
the power spectrum is given by 
\begin{equation}
    \langle B_{\rm s}^2\rangle = \langle B_{\rm s,pos}^2 + B_{\rm s,neg}^2 + 2B_{\rm s,pos}B_{\rm s,neg}\rangle  = B_0^2\langle f_{\rm s}^2 v_{\rm s,pos}^2 + f_{\rm s}^2 v_{\rm s,neg}^2 - 2f_{\rm s}^2 v_{\rm s,pos} v_{\rm s, neg} \rangle \ .
\end{equation}
Note the sign difference in the mixing terms. 

Lastly, there are some variables that are extremely well correlated, namely the density fluctuation $\delta\rho$ and the magnetic field fluctuation in the parallel direction, $\delta B_{\parallel}$. One can see that they should be negatively correlated by directly looking at the equations. Interestingly, this correlation is kept very well while other correlations start to break at higher injection rates, a hint that this correlation is more robust than other linear equations of motion. We will discuss this further in section~\ref{sec:injection_rates}.

\subsection{CMB polarizations}
\begin{figure}[ht]
    \centering
    \includegraphics[width=0.8\textwidth]{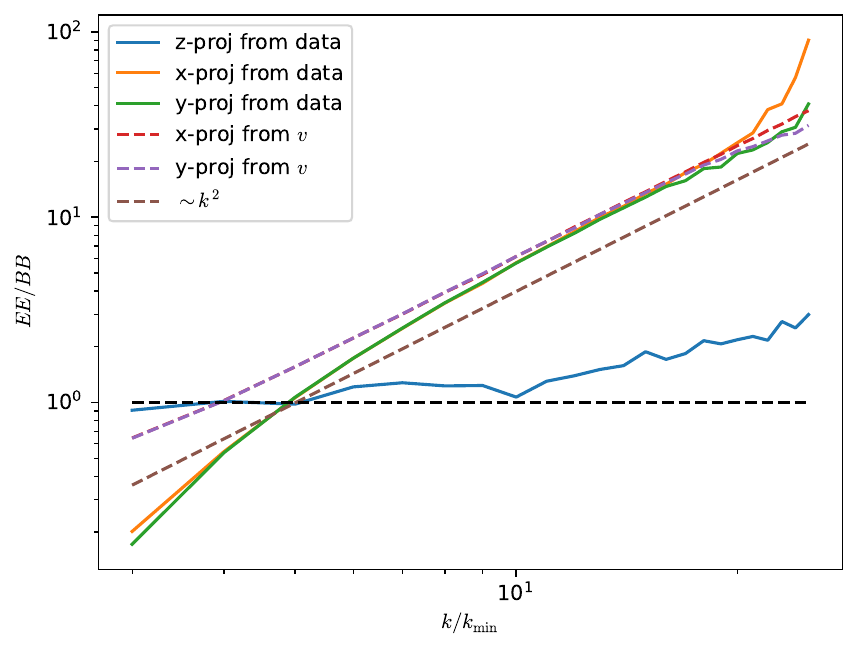}
    \caption{$EE/BB$ ratio directly measured from the cubes and reconstructed from the velocity fields via linear equations, at injection rate $\epsilon = 0.001$. We see that they both show a trend of $\sim k^{2}$ increase and agree with each other at around $k\sim 10$. The calculated $E$ and $B$ modes are both zero for the z projection, so we omit the z-proj from $v$ line. The measured value comes from deviations from linear relations, and is included for completeness.}
    \label{fig:EB_ratio}
\end{figure}

Similar to the density and magnetic field fluctuations, using a simple dust radiation model \cite{Kritsuk:2017aab}, we can calculate the $E$ and $B$ mode polarizations coming from dust radiation. In this quasi-linear regime, using again the results from \cite{Caldwell:2016xkd,Kandel:2017xjx}, we can write
\begin{align}
    E \propto& -\sin 2\theta \frac{\sin \psi}{\sin \alpha} \frac{\delta B_{\rm a} }{B_0} + \frac{\sin^2 \theta \left( -2\sin^2 \psi (1-2\sin^2\alpha) - 2\sin^2 \alpha  \right)}{\sin\alpha} \frac{\delta B_{\rm s/f}}{B_0} + \sin^2 \theta \cos 2 \psi \frac{\delta \rho}{\rho_0}, \\
    B \propto& -\sin 2\theta \frac{\cos \psi}{\sin \alpha} \frac{\delta B_{\rm a} }{B_0} + \frac{2\sin^2 \theta \sin \psi \cos \psi(1-2\sin^2 \alpha)}{\sin\alpha} \frac{\delta B_{\rm s/f}}{B_0} + \sin^2 \theta \sin 2 \psi \frac{\delta \rho}{\rho_0} \ .
\end{align}
For this setup, the line of sight is along the $z$ axis, the wavevector in the $x-y$ plane is given by $\vec k = (\cos\psi,\sin\psi,0)|k|$, the mean magnetic field is set to be in the $x-z$ plane so that $\vec B_0 = (\sin \theta ,0, \cos \theta )B_0$, and the angle between the magnetic field and the wavevector is $\alpha$, which can be calculated by $\cos \alpha = \cos \psi \sin \theta $. Specializing to our case, we only project parallel or perpendicular to the mean magnetic field, so that $\theta = 0, \pi/2$. The case $\theta = 0$ gives zero contributions to the $E$ and $B$ modes, and one can actually check in the data that the polarization spectra in the parallel projection are smaller than those in the perpendicular projection by several orders of magnitude. We therefore focus on the perpendicular projection $\theta = \pi/2$, and take $\alpha = \psi$. Further dropping the fast mode contribution, 
\begin{align}\label{eq:E_B_mode_lin}
    E_{\perp} \propto&  -4 \sin \alpha \cos^2\alpha \frac{\delta B_{\rm s}}{B_0} + \cos 2 \alpha \frac{\delta \rho_{\rm s}}{\rho_0}, \\
    B_{\perp} \propto& \ \ 2 \cos \alpha \cos 2\alpha \frac{\delta B_{\rm s}}{B_0} + \sin 2 \alpha \frac{\delta \rho_{\rm s}}{\rho_0} \ .
\end{align}
Note that there is no contribution from Alfv\'en modes. We then use the density and magnetic field fluctuations calculated from velocity fields to compute the $E$ and $B$ mode polarization spectra through these equations, and compare with direct measurements from the simulations. For our purpose, we focus on the ratio between the $E$ mode and $B$ mode spectra. As shown in Fig.~\ref{fig:EB_ratio}, the calculated and measured values match well around $k/k_{\rm min} \sim 10$. 

\begin{figure}
     \centering
     \begin{subfigure}[b]{0.49\textwidth}
         \centering
         \includegraphics[width=\textwidth]{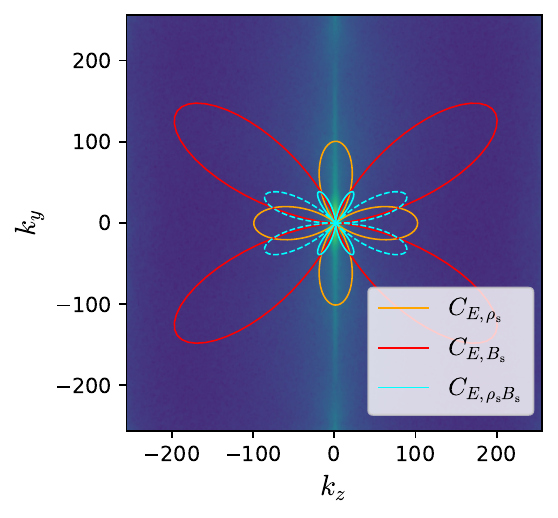}
     \end{subfigure}
     \hfill
     \begin{subfigure}[b]{0.49\textwidth}
         \centering
         \includegraphics[width=\textwidth]{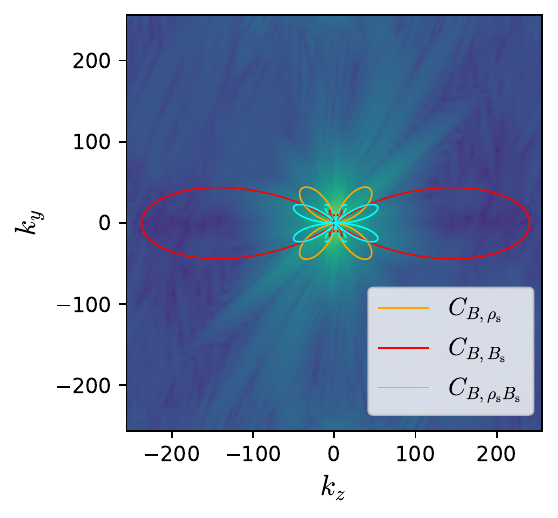}
     \end{subfigure}
     
    \caption{Polar plots of the coefficients in front of the power spectra to convert from $B$ and $\rho$ fluctuations to polarization spectra. On the left we have the coefficients for $E^2$ on top of the the $\delta \rho$ spectrum at $\epsilon = 0.001$, and on the right we have the coefficients for $B^2$ on top of the $\delta \rho$ spectrum at $\epsilon = 0.1$. Dashed lines mean the coefficients are negative at this angle. We see that the $E$ mode spectra probes the dominating part when the 2D spectrum is highly anisotropic, and therefore $\delta \rho$ contribution dominates, while the $B$ mode spectra probes the less dominating regions. We also see the spectrum becomes less anisotropic, with rays pointing at various different directions instead of only perpendicular to $\vec B_0$ as $\epsilon$ increases.}
    \label{fig:EE_BB_coeff}
\end{figure}

A major signature of the result is the $\sim k^2$ scaling of the $E$ to $B$ ratio. Note that this differs significantly from the behavior seen in the {\em Planck} data where the $E$ to $B$ ratio is roughly flat \cite{Planck:2014dmk}. We attribute this behavior to the anisotropy of the 2D spectra of the MHD fields. To go into the details, let us focus on the slow mode behavior in the equations for the power spectra. Suppose the $E$ and $B$ modes spectra are calculated by 
\begin{align}
     E^2  =& C_{E,\rho_{\rm s}}P_{\rho_{\rm s}} + C_{E,B_{\rm s}} P_{B_{\rm s}} + C_{E,\rho_{\rm s}  B_{\rm s}}P_{\rho_{\rm s} B_{\rm s}}  \ , \\
     B^2  =& C_{B,\rho_{\rm s}}P_{\rho_{\rm s}} + C_{B,B_{\rm s}} P_{B_{\rm s}} + C_{B,\rho_{\rm s}  B_{\rm s}}P_{\rho_{\rm s} B_{\rm s}}   \ ,\label{eq:E_B_spec_lin}
\end{align}
where $P_{\rho_{\rm s}/ B_{\rm s} / \rho_{\rm s} B_{\rm s}}$ are the power spectra of the $\rho$, $\vec B$, and the cross spectrum of the two fields in slow mode, respectively. Plotting the 2D spectra on the $k_x = 0 $ or $k_y = 0$ plane, we see an apparent anisotropic behavior, as a function of the angle between $\vec k$ and $\vec B_0$. Now plot the $C$ coefficients in the perpendicular projection as a function of the angle between $\vec k$ and $\vec B_0$, as in Fig.~\ref{fig:EE_BB_coeff}, we see that they screen different parts of the 2D spectra. Specifically, the $E$ mode coefficient $C_{E,\rho_{\rm s}}$ emphasizes the part $\vec k \perp \vec B_0 $\footnote{It also probes the $\vec k \parallel \vec B_0$ part, but this direction is sub-dominant compared to all other directions.},  where the power is the highest, while the coefficients for $B$ modes go to zero for $\vec k \perp \vec B_0$, and are the highest along the diagonals. Hence the increasing $E$ to $B$ ratio reflects the increasing anisotropy of turbulence toward smaller scales. The increasing anisotropy was predicted based on the critical balance conjecture in \cite{1995ApJ...438..763G}. We will see later how this behavior is mitigated by compressibility effects as the energy injection rates grow. 

We also want to note that due to the behavior of the $C$ coefficients mentioned above, the $E$ mode is dominated by the spectrum $P_{\rho_{\rm s}}$, while for the $B$ mode, all three terms are important. That is, linearly the $E$ mode is a direct reflection of density fluctuations. On the other hand, if we were to take out the cross term in the $B$ mode, we would miss important contributions to the $B$ mode spectrum, leading to a flawed prediction for the $E$ to $B$ ratio.

\section{Increasing injection rates}
\begin{figure}
     \centering
     \begin{subfigure}[b]{1\textwidth}
         \centering
         \includegraphics[width=\textwidth]{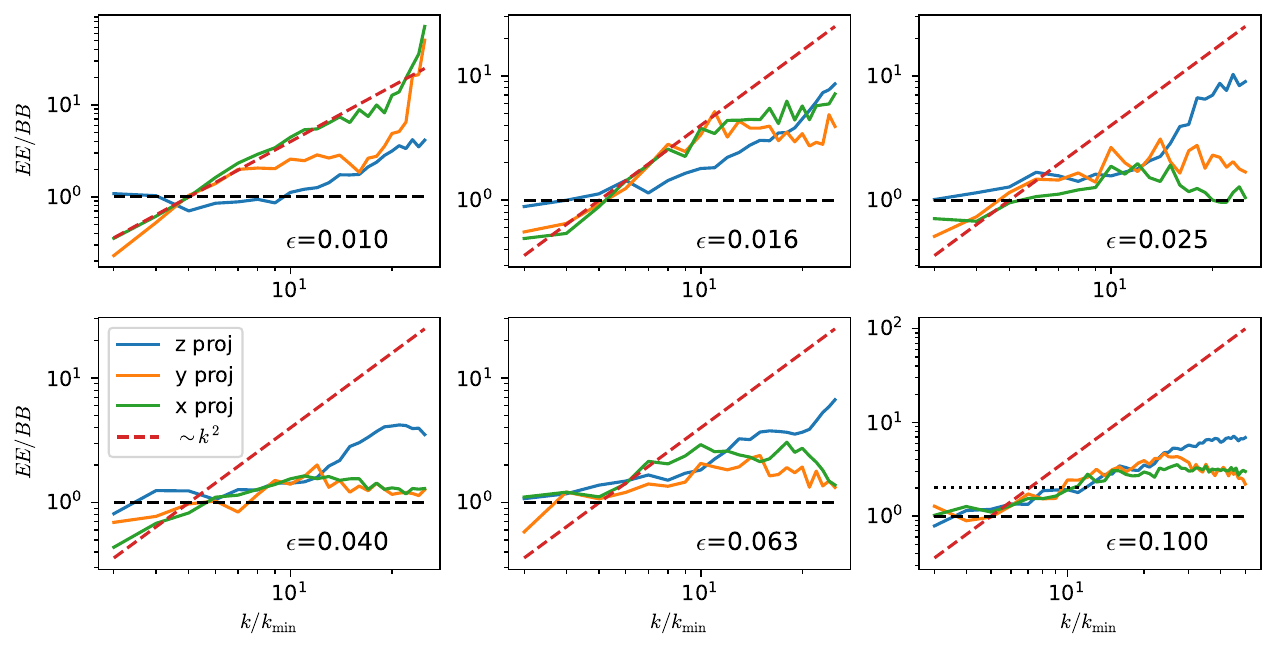}
     \end{subfigure}
     \vfill
     \begin{subfigure}[b]{1\textwidth}
         \centering
         \includegraphics[width=\textwidth]{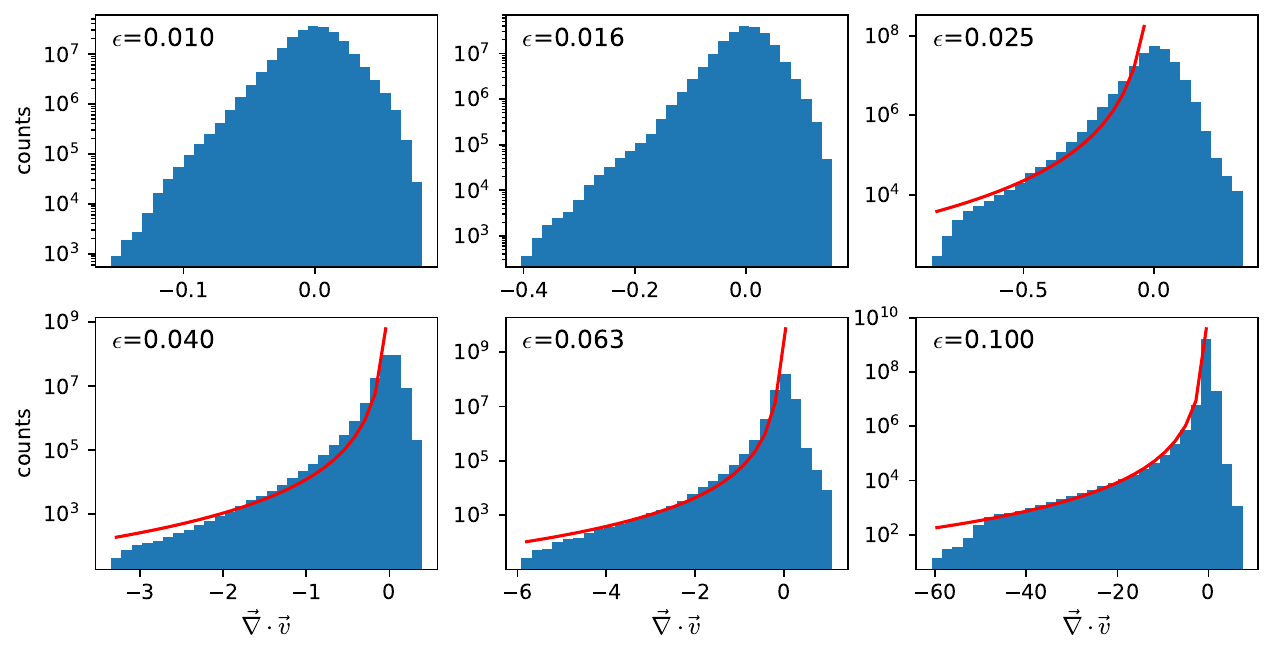}
     \end{subfigure}
     
    \caption{Flattening of the $EE/BB$ ratio and the behavior of the $\vec \nabla \cdot \vec v$ distribution with increasing energy injection rate $\epsilon$. We see that with increasing injection rates the $EE/BB$ ratio gradually flattens to a value larger than 1 (we use dashed black line to indicate the value 1, and in the $\epsilon = 0.1$ case we use dotted black line to represent the value 2). The $\vec \nabla \cdot \vec v$ distribution develops a negative tail with a power law behavior with exponenet $\sim -7/2$, marked by the red lines, hinting the formation of shocks.}
    \label{fig:increasing_rates}
\end{figure}

{\label{sec:injection_rates}}
Now let us check the behavior of the system with increasing energy injection rates. In our previous quasi-linear analysis, we have used an injection rate $\epsilon = 0.001$. To see what happens at higher injection rates, we carried out several simulations at $256^3$ grid resolution at $\epsilon = 0.010, 0.016, 0.025, 0.040, 0.063$, and another $512^3$ simulation at $\epsilon = 0.1$. Quite interestingly, there is a clear qualitative transition happening around $\epsilon \sim 0.01$ to $0.02$, corresponding to turbulence Mach number $M_{\rm t}\sim0.4$. Below this injection rate, the distribution of the divergence $\vec \nabla \cdot \vec v$ is not too far from a Gaussian, without a prominent tail, the $EE/BB$ ratio has a power law behavior of $\sim k^{2}$, and the 2D spectra of all the fields are highly anisotropic. Above this injection rate, we see the development of a large negative tail of $\vec \nabla \cdot \vec v$, with a distribution close to a power law with exponent $\sim -7/2$, hinting that we start to see shocks forming, and the system approaches Burgers-like regime \cite{2025arXiv250418174K,2024PhRvL.133t7201K}. At the same time, the $EE/BB \sim k^2$ scaling gradually flattens to a much shallower behavior, while the 2D spectra get less anisotropic. See Fig.~\ref{fig:EE_BB_coeff} for the change in 2D spectra and Fig.~\ref{fig:increasing_rates} for the flattening of $EE/BB$ ratio with increasing injection rates and the change of the distribution of $\vec \nabla \cdot \vec v$. At this level of injection rate, the distribution of the direction of the local magnetic field is not disperse enough to explain the isotropization of the 2D spectra. Combining all the features, we propose that the formation of shocks is key to the isotropization of the 2D spectra and the flattening of the $EE/BB$ ratio.

One point worth discussing is that the $EE/BB$ ratio flattens to a number that is larger than 1 consistently. We can deduce from the linear relations~(\ref{eq:E_B_mode_lin})-(\ref{eq:E_B_spec_lin}), by simply integrating over the angle $\alpha$, that for totally isotropic spectra, a linear relation between the polarization modes and the MHD fluctuations leads to $EE/BB \lesssim 1$. Therefore there have to be some mechanisms that boost the power of $E$ mode with respect to the $B$ mode. In multiphase turbulence simulations, the high density cold neutral medium, where shocks are prominent but hydrodynamic-like (i.e. highly isotropic, with little influence from the magnetic field), indeed generates an $EE/BB$ ratio that is flat, but closer to unity. In contrast, in the lower density phases, regardless of whether cold or warm, where the system is more MHD-like (i.e. dynamics are affected by magnetic fields and structures retain some degree of anisotropy rather than being highly isotropic), the $EE/BB$ ratio is close to 2~\cite{Flauger2018}. In this sense, not only do we need shocks to achieve a flat $EE/BB$ ratio, but we also need {\it MHD shocks} to reproduce the observed value. Physically, this is because anisotropic properties of MHD shocks lead to more power in $E$ modes than in $B$ modes, similar to what happens in filaments (see, e.g., ~\cite{Huffenberger_2020} for an analytical study). We also know from supersonic isothermal simulations that after masking extreme high density voxels, the resulting $EE/BB$ ratio is close to $2$~\cite{Flauger2018}. That is, the presence of multiple thermal phases is not necessary for a higher $EE/BB$ ratio. In addition, there could be other nonlinear mechanisms that lead to this result. For example, the fluctuations of the local mean magnetic field could be large and lead to problems in mode decomposition and projections in the linear equations. We leave investigations of these nonlinear effects to future work. 

\begin{figure}[ht]
    \centering
    \includegraphics[width=0.8\textwidth]{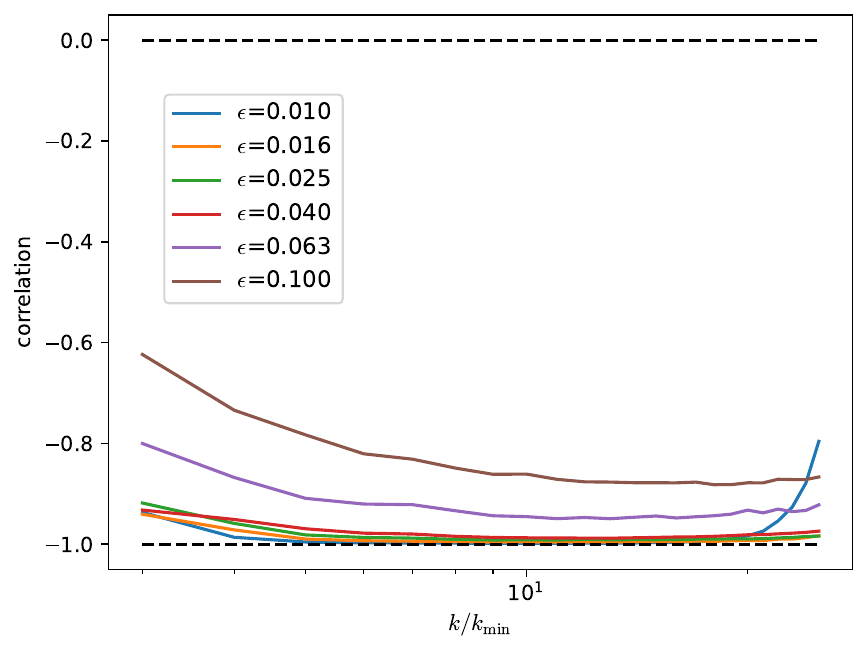}
    \caption{Evolution of the correlation between $\delta B_\parallel$ and $\delta \rho$ with increasing $\epsilon$. We see that the correlation stays close to $-1$ until the injection rate is significantly higher than the transition value $\sim 0.2$.}
    \label{fig:BparaRhoCorr}
\end{figure}

Lastly, as mentioned before, the correlation between $\delta \rho $ and $\delta B_{\parallel}$ remains extremely good, even with increasing injection rates. We showcase this in Fig.~\ref{fig:BparaRhoCorr}. At leading order, we could explain the anti-correlation by looking directly at the linear equations. Since the power in the spectra is mainly contained in the perpendicular direction, we can analyze the ratio between $\delta \rho_{\rm s} $ and $\delta B_{\rm s,\parallel}$\footnote{At leading order $\delta B_{\rm s, \parallel} = \delta B_{\parallel}$ since $\delta \vec B_{\rm a} \perp \vec B_0$.} at the limit $\alpha \rightarrow \pi/2$,
\begin{equation}
    \lim_{\alpha \rightarrow \pi/2} \frac{\delta \rho_{\rm s}/\rho_0}{\delta \vec B_{\rm s,\parallel}/B_0} = - \lim_{\alpha \rightarrow \pi/2}\frac{\cos^2 \alpha + \zeta_{\rm s} \sin^2\alpha}{\zeta_{\rm s} \sin^2 \alpha} = -\frac{2}{\beta} \ .
\end{equation}
An interesting consequence is that the total pressure in the isothermal case, given by $B^2/8\pi + \rho c_{\rm s}^2$, is constant at leading order. It corresponds to a balancing between the incompressible and compressible parts in the language of nearly incompressible MHD \cite{zank93}. Such a balancing behavior has been observed in nature, for example in the solar wind, and it has been suggested in the literature that this is a signature of the dominance of slow and Alfv\'en modes \cite{2012ApJ...753L..19H}. A similar discussion can also be found in \cite{2001ApJ...562..279L}. What is more interesting is that, at an injection rate $\epsilon \sim 0.1$, where the linear waves become unstable and start to break, most other correlations are already broken badly, while this one persists at around $-0.8$. The formation of shocks does not destroy it. That is, this relation is less prone to increasing injection rates and survives the onset of nonlinearities. It is therefore something that goes beyond simple linear relations.  Physically, since both slow magnetosonic waves and slow shocks have the property that higher density leads to lower magnetic field strength (see e.g. \cite{Biskamp_2003}), it is not very surprising that this property survives the onset of the slow wave-breaking regime. Lastly, at even higher injection rates, for example for $\epsilon \ge 1$, this correlation indeed breaks, as density fluctuations can become too high for magnetic pressure to balance out. 
 
\section{Conclusions and perspective} 
We have shown that in the quasi-linear regime, the linearized MHD equations together with the large scale advection effects predict the behavior of the simulations to good precision, and a transition from quasi-linear to MHD shock-dominated regime leads to the flattening of $EE/BB$ ratio to a value that is consistently larger than 1. The flattening is related to isotropization of the 2D spectra as a consequence of formation of shocks. We see from the behavior of divergence of velocity that shocks start to become significant exactly where the $EE/BB$ ratio flattens. 
Moreover, MHD shocks rather than isotropic hydrodynamic shocks further boost the $EE/BB$ ratio to be $> 1$. The density fluctuations maintain a very good negative correlation with the magnetic fluctuations parallel to the mean magnetic field, confirming the dominance of slow magnetosonic waves. As a result, to correctly understand the observed flat $EE/BB$ ratio, one should go beyond linearized MHD equations, and take into account the effects of MHD shocks and other possible non-linear effects. 

The simulations we use in this work are near-isothermal single thermal phase simulations. It would be interesting to investigate the role shocks play in multiphase simulations of ISM turbulence. In addition, fast modes start to be significant at some point, which may lead to new phenomenological behaviors. Lastly, there could be some principles that are more general than the linear equations of motion that protect the pressure balance behavior at decently high energy injection rates. It would be interesting to pin down these principles. We leave these investigations to future work.

\paragraph{Acknowledgments}
The authors thank Andrey Beresnyak, Ka Wai Ho, and Ka Ho Yuen for illuminating discussions. GS also thanks Ashley Bransgrove and Weiyi Li for their priceless help when learning MHD and turbulence. AK and RF were supported in part by the NASA Grant No. 80NSSC22K0724; RF and GS were supported by the US~Department of Energy under grant~\mbox{DE-SC0009919}. GS is also supported in part by the National Natural Science Foundation of China under Grant No. 12547101. Computational and storage resources were provided by the ACCESS program (MCA07S014) and by the NASA High-End Computing (HEC) Program through the NASA Advanced Supercomputing (NAS) Division at Ames Research Center.

\bibliography{main,adpdis}{}

\bibliographystyle{utphys} 

\end{document}